\documentclass{ragtime}

\usepackage{times}
\usepackage{mathrsfs}

\title[Notes on generic rotating charged regular black holes]{Note on the character of the generic rotating charged regular black holes in general\\ relativity coupled to nonlinear electrodynamics}

\author[B. Toshmatov, % Run. head authors: separate names with commas,
        Z. Stuchl\'{i}k]{Bobir Toshmatov\at{1,a} Zden\v{e}k Stuchl\'{i}k\at{1,b} Bobomurat Ahmedov\at{2,c} \\ \ins{1} Institute of Physics and Research Centre of Theoretical Physics and Astrophysics,\splitins[1] Faculty of Philosophy \& Science, Silesian University in Opava, \splitins[1] Bezru\v{c}ovo n\'{a}m\v{e}st\'{i} 13,  CZ-74601 Opava, Czech Republic\\ \ins{2} Ulugh Beg Astronomical Institute, Astronomicheskaya 33, Tashkent 100052, Uzbekistan\\
        \ins{a}\Email{bobir.toshmatov@fpf.slu.cz}
        \ins{b}\Email{zdenek.stuchlik@fpf.slu.cz}\\
        \ins{c}\Email{ahmedov@astrin.uz}}

\graphicspath{{pic/}}

\begin{document}

\begin{abstract}
We demonstrate that the generic charged rotating regular black hole solutions of general relativity coupled to non-linear electrodynamics, obtained by using the alternate Newman-Janis algorithm, introduces only small (on level $10^{-2}$) inconsistency in the behaviour of the electrodynamics Lagrangian. This approves application of these analytic and simple solutions as astrophysically relevant, sufficiently precise approximate solutions describing rotating regular black holes.
\end{abstract}


\section{Introduction}

Starting by the Bardeen geometry~\citep{Bar}, regular black hole solutions attract extended attention till present times. A special focus is devoted to the regular black hole solutions in general relativity combined with the nonlinear electrodynamic models~\citep{Hay,ABG,Neves}. The spherically symmetric solutions were studied in variety of works~\citep{Bronnikov,s-s}, recently generic black hole solutions of general relativity coupled to the Born-Infeld electrodynamics were introduced in~\citep{f-w} that could cover many of the previously introduced solutions. The rotating regular black hole solutions are usually generated by the Newman-Janis algorithm (NJA)~\citep{n-j}, or by its modification~\citep{a-a}. The modified NJA was applied in the case of the generic rotating charged black holes in~\citep{t-s-a}. However, an inconsistency related to the behaviour of the Lagrangian of the nonlinear electrodynamics has been noticed in~\citep{r-j}. Here we shortly estimate this inconsistency and its implication on the relevance of the generic rotating regular solutions presented in~\citep{t-s-a}.

\section{Inconsistency of the generic rotating black hole solution in (Toshmatov et al.,~2017)}

In the paper~\citep{t-s-a} we had recently obtained the following solution which is one of the possible candidates for the rotating regular black hole solution of general relativity coupled to nonlinear electrodynamic field, converting the spherically symmetric regular black hole solution~\citep{f-w} by using the alternate NJA. The spacetime geometry of this solution reads
\begin{eqnarray}\label{rotating2}
ds^2&=&-\left(1-\frac{2\rho r}{\Sigma}\right)dt^2+\frac{\Sigma}{\Delta}dr^2-2a\sin^2\theta\frac{2\rho r}{\Sigma}d\phi dt\nonumber\\
&&+\Sigma d\theta^2+\sin^2\theta\frac{(r^2+a^2)^2-a^2\Delta\sin^2\theta}{\Sigma}d\phi^2\ ,
\end{eqnarray}
with
\begin{eqnarray}\label{notations}
&&\Sigma=r^2+a^2\cos^2\theta, \qquad 2\rho=r(1-f),\nonumber\\
&&\Delta=r^2f+a^2=r^2-2\rho r+a^2\ .
\end{eqnarray}
where $\rho$ is the mass function which depends on radial coordinate and electrodynamic field parameters and $f$ is determined by the spherically symmetric class of solutions of the theory. These solutions can take the form of the Bardeen-like, Hayward-like, and Maxwell-like spacetimes. The generic solution~(\ref{rotating2}) had been obtained analytically and it satisfies the Einstein equations in the tetrad frame -- see~\citep{t-s-a} for details. However, the NJA does not always lead to true precise solutions of the whole set of field equations of the theory under consideration, i.e, the energy-momentum tensor of the rotating regular black hole solution generated by the NJA is not exactly fulfilling the equations of nonlinear electrodynamics in some cases, as explicitly demonstrated in~\citep{t-s-a,r-j}.

\begin{figure}
\includegraphics[width=0.8\textwidth]{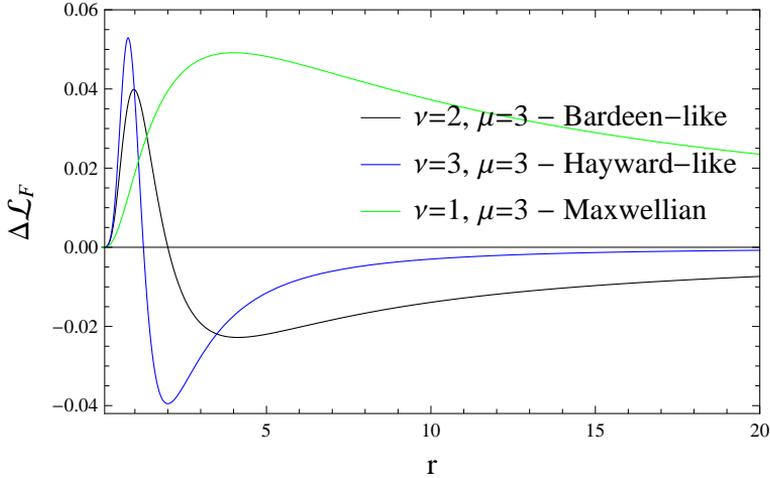}
\caption{\label{fig1} Plot illustrates a change of difference (estimation of inconsistency) $\Delta\mathscr{L}_F=\mathscr{L}_F-\partial\mathscr{L}/\partial F$ with the radial coordinate $r$ at the equatorial plane of the rotating regular black holes presented by~\cite{t-s-a} for $q=1$, $a=0.5$.}
\end{figure}

In the case of the nonrotating black holes with total magnetic charge $Q_m$, considered in nonlinear electrodynamics, the Lagrangian density of the nonlinear electrodynamic field is defined in terms of $\rho$ as~\citep{f-w}
\begin{eqnarray}\label{ned}
\mathscr{L}=\frac{4\rho'}{r^2}\ ,\qquad
\mathscr{L}_F\equiv\frac{\partial\mathscr{L}}{\partial F}=\frac{r^2(2\rho'-r\rho'')}{2Q_m^2}\ .
\end{eqnarray}
where the electromagnetic field strength is $F=2Q_{m}^2/r^4$. As a rule, introduction of the rotation parameter by the NJA must change the form of the Lagrangian density~(\ref{ned}), and the gauge of the field as well. The new gauge can be easily found (for details -- see~\citep{t-s-a}). However, we cannot apply the NJA directly to~(\ref{ned}). The only way to find the Lagrangian density of the rotating black hole in nonlinear electrodynamics, is to solve the Einstein field equations, $G_{\mu\nu}=T_{\mu\nu}$, with respect to $\mathscr{L}$ and $\mathscr{L}_F$ in the background~(\ref{rotating2}). Here, the energy-momentum tensor of the nonlinear electrodynamic field is given by
\begin{eqnarray}\label{emt}
T_{\mu\nu}=2\left(\mathscr{L}_F F_\mu^\alpha F_{\nu\alpha}- \frac{1}{4}g_{\mu\nu}\mathscr{L}\right)\ ,
\end{eqnarray}
Thus, the Einstein equations give five independent equations with two unknowns $\mathscr{L}$ and $\mathscr{L}_F$, which cannot be solved simultaneously. Therefore, in the paper~\citep{t-s-a} we had not solved the whole set of equations, instead, we solved just three of them simultaneously, and obtained the expressions for $\mathscr{L}$ and $\mathscr{L}_F$ in the form
\begin{eqnarray}
\mathscr{L}=&&\frac{r^2\left(15a^4-8a^2r^2+8r^4+4a^2(5a^2-2r^2)\cos2\theta+5 a^4\cos4\theta\right)\rho'}{2\Sigma^4}\nonumber\\ &&+\frac{8a^2r^3\cos^2\theta\rho''}{\Sigma^3},\label{lagrangian}\\
\mathscr{L}_F=&&\frac{2(r^2-a^2\cos^2\theta)\rho'-r\Sigma\rho''}{2Q_{m}^2},\label{lagrangian2}
\end{eqnarray}
where the electromagnetic field strength $F$ in the rotating case takes the following form~\citep{t-s-a}
\begin{eqnarray}\label{squaredF}
F=\frac{Q_{m}^2[a^4(3-\cos4\theta)+4(6a^2r^2+2r^4 +a^2(a^2-6r^2)\cos2\theta)]}{4\Sigma^4}.
\end{eqnarray}
In the nonrotating limit, $a=0$, we recover the expressions~(\ref{ned}). Obviously, these obtained expressions of $\mathscr{L}$, Eq.~(\ref{lagrangian}), and $\mathscr{L}_F$, Eq.~(\ref{lagrangian2}), do not exactly satisfy the remaining two equations. Therefore, as it has been shown in the paper~\citep{r-j}, the total derivative of $\mathscr{L}$ with respect to $F$ is not equal to $\mathscr{L}_F$. The difference can be written as
\begin{eqnarray}\label{eq1}
\Delta\mathscr{L}_F=\mathscr{L}_F-\frac{\partial\mathscr{L}}{\partial F}\equiv\mathscr{L}_F-\frac{\partial\mathscr{L}}{\partial r}\frac{\partial r}{\partial F}-\frac{\partial\mathscr{L}}{\partial\theta}\frac{\partial\theta}{\partial F}\neq0.
\end{eqnarray}
However, we are able to demonstrate explicitly that the values of $\mathscr{L}_F$ and $\partial\mathscr{L}/\partial F$ are very close and thus, the remaining equations can be considered as being approximately fulfilled, as the value of $\Delta\mathscr{L}_F$ is close to zero. In Fig.~\ref{fig1} we present the radial profile of this difference, $\Delta\mathscr{L}_F$, for the typical values of the Bardeen-like, Hayward-like and Maxwell-like rotating regular black holes.
One can see from Fig.~\ref{fig1} that the inconsistency in the rotating regular black hole solution is very small, on the level of $10^{-2}$. Therefore, our results can be considered in the same spirit as many results obtained in relation to the so called dirty Kerr-like spacetime metrics, where the relation to the mass stress energy tensor is not considered at all, but the results are considered as relevant for estimations of the astrophysical phenomena related to uncharged matter. Of course, in the case of the behaviour of charged matter in the charged rotating regular black hole backgrounds, we have to be very careful due to the small inconsistency related to the electrodynamic part of the theory.

%\subsection{Comments on the alternative method of constructing rotating solution~\cite{d-g1}}

%In the paper~\cite{r-j}, the authors suggested to use another method, as proposed by Dymnikova and Galaktionov~\cite{d-g1} for obtaining the electrically charged rotating regular black hole solution in general relativity minimally coupled to the nonlinear electrodynamic field. To obtain rotating solution from the spherically symmetric one they used the G\"{u}rses-G\"{u}rsey algorithm instead of the NJA. However, finding the exact solution of the rotating regular black hole solution in general relativity minimally coupled to the nonlinear electrodynamic field analytically is then very complex process, since the relevant equations are highly nonlinear in terms of the mass function. Moreover, finding the magnetically charged solution requires to solve even more complicated nonlinear differential equations with respect to radial and azimuthal coordinates. Therefore, this method also fails in attempt to obtain analytical rotating regular both electrically and magnetically charged black hole solutions. Clearly, the generic charged rotating regular black hole solution of general relativity combined with nonlinear electrodynamics can be considered as astrophysically relevant approximative solutions, having the advantage of being analytic solutions with simple Kerr-like form.

\section{Conclusion}

In general situations, for generic regular charged black hole spherically symmetric solutions of general relativity combined with the nonlinear electrodynamics, it is very difficult to find corresponding rotating black hole solutions in an analytic and fully precise form. Probably, the exact and consistent solutions could be constructed only by numerical procedures. However, the generic rotating charged black hole solutions obtained by using the alternate NJA in~\citep{t-s-a} are analytic and simple solutions that are precise enough for exploring such solutions in astrophysical situations involving uncharged matter.

\ack
This work is supported by the internal student grant of the Silesian University (Grant No. SGS/14/2016) and the Albert Einstein Centre for Gravitation and Astrophysics under the Czech Science Foundation (Grant No.~14-37086G).

%\label{lastpage}

\bibliography{reference}

\end{document}